\pdfoutput=1
\documentclass[11pt,twoside,a4paper,cmspaper,final,collab]{cms-tdr}
\def\svnVersion{2986488}\def\svnDate{2026/09/12}\def\cmsCernNoTag{CERN-EP-2026-023}\def\cmsCernDate{\today}\def\cmsMessage{Published in Physical Review Letters as \href{http://dx.doi.org/10.1103/dv54-jq8p}{\doi{10.1103/dv54-jq8p}.}}
\begin{document}\cmsNoteHeader{BPH-24-001}

\newcommand{\twomutwoe}{\ensuremath{\PGh\to\PGmp\PGmm\Pep\Pem}\xspace}
\newcommand{\twomuG}{\ensuremath{\PGh\to\PGmp\PGmm\PGg}\xspace}
\newcommand{\Btwomutwoe}{\ensuremath{\mathcal{B}(\twomutwoe)}\xspace}
\newcommand{\BtwomuG}{\ensuremath{\mathcal{B}(\twomuG)}\xspace}
\newcommand{\etatomumu}{\ensuremath{\PGh\to\PGmp\PGmm}\xspace}
\newcommand{\foure}{\ensuremath{\PGh\to\Pep\Pem\Pep\Pem}\xspace}
\newcommand{\emmg}{\ensuremath{\epsilon_{2\PGm\PGg}}\xspace}
\newcommand{\emmee}{\ensuremath{\epsilon_{2\PGm 2\Pe}}\xspace}
\newcommand{\Nmmee}{\ensuremath{N_{2\PGm 2\Pe}}\xspace}
\newcommand{\Nmmg}{\ensuremath{N_{2\PGm\PGg}}\xspace}
\newcommand{\chisq}{\ensuremath{\chi^2}\xspace}

\cmsNoteHeader{BPH-24-001}
\title{Observation of the rare decay \texorpdfstring{$\twomutwoe$}{eta to mu+mu-e+e-}}
\date{\today}
\abstract{
A first observation of the rare decay $\twomutwoe$ is reported by the CMS Collaboration at the CERN LHC. The result is based on a proton-proton collision data sample at $\sqrt{s} = 13.6\TeV$ corresponding to an integrated luminosity of 38.0\fbinv, acquired in 2022 using a high-rate dimuon trigger. Using the $\twomuG$ decay channel for normalization, the branching fraction is measured to be $\Btwomutwoe = (2.4 \pm 0.8)\times 10^{-6}$, with the uncertainty including statistical and systematic sources as well as the $\BtwomuG$ uncertainty. This result is about a factor of three smaller than the existing limit, and is consistent with recent theoretical predictions.
}
\hypersetup{%
pdfauthor={CMS Collaboration},%
pdftitle={Observation of the rare decay eta to mu+mu-e+e-},%
pdfsubject={CMS},%
pdfkeywords={CMS, eta meson, data-parking, rare decays, experimental results}} 

\maketitle 
The \PGh and \PGhpr mesons are light, flavor- and charge-neutral, spin-0 particles, with masses of 547.86(2) and 957.78(6)\MeV, respectively~\cite{pdg}. 
The measurement of the rate for rare decays of these particles is motivated by several considerations: they assist in theoretical computations of the light-by-light hadronic contributions to the muon anomalous magnetic moment~\cite{Knecht}, enable novel tests of lepton flavor universality in particle physics~\cite{cms4mu}, and are sensitive to beyond-standard-model theories that predict deviations in these rates~\cite{Gan,Goudzovski}.

\begin{figure}[ht]
\centering
\includegraphics[width=0.5\textwidth]{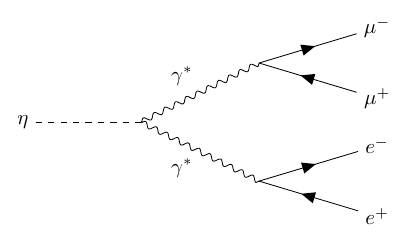}
\caption{Leading order standard model Feynman diagram for the decay \twomutwoe.}
\label{fig:feynman}   
\end{figure}

Single- and double-Dalitz decays of the \PGh and \PGhpr mesons to purely leptonic final states proceed via chiral anomalies with one or two virtual photons, respectively, and their widths depend on theoretically unknown transition form factors that model the low-energy quantum chromodynamic interactions. 
Theoretical predictions for these decays have been made based on chiral perturbation theory~\cite{Bijnens}, its extensions such as Vector Meson Dominance~(VMD)~\cite{VMD0,VMD1,Petri} or the light-front quark model~\cite{Lih}, and further extensions of VMD using a data-driven approach~\cite{Escribano,Kampf}. 
These data-driven methods are the ones relied upon for muon anomalous magnetic moment calculations because of their assumed accuracy~\cite{gm2calc,gm2calc2}.
Some of these decays have been experimentally observed and, thus far, none of the measured branching fractions~(BFs) deviates significantly from the data-driven theoretical predictions (presented in Ref.~\cite{Gan} and references therein).
The modes that have been observed include \etatomumu~\cite{eta2muOg,eta2muNew}, \foure~\cite{eta4e}, $\PGhpr\to\Pep\Pem\Pep\Pem$~\cite{etap4e}, and most recently, by CMS, $\PGh\to\PGmp\PGmm\PGmp\PGmm$~\cite{cms4mu}.
However, several decay modes have yet to be observed, as is the case of the double-Dalitz decay \twomutwoe, the first observation of which is reported here.
The leading order standard model Feynman diagram for this decay is shown in Fig.~\ref{fig:feynman}.
Using a special high-rate dimuon trigger technique developed in the CMS experiment, an improvement of about a factor of three in the experimental sensitivity of this decay mode's BF is achieved when compared to the previous upper limit of $6.88 \times 10^{-6}$ at the 90\% confidence level~\cite{besIII}.

The data used for this observation are from $\sqrt{s} = 13.6\TeV$ proton-proton collisions in the CERN LHC, collected in 2022 by the CMS experiment, corresponding to an integrated luminosity of 38.0\fbinv.
Tabulated results are provided in the HEPData record for this analysis~\cite{hepdata}.
To measure the BF for the \twomutwoe signal, the \twomuG mode, which has a measured BF of $(3.1 \pm 0.4) \times 10^{-4}$~\cite{pdg}, is used as reference. 
The photon in the final state subsequently converts to an electron-positron pair in the inner region of the CMS detector, meaning the signal and reference channels have the same final-state particles.
The BF can be calculated using
\begin{equation}
    \frac{\mathcal{B}_{2\PGm 2\Pe}}{\mathcal{B}_{2\PGm\PGg}} = \frac{\Nmmee}{\Nmmg}\frac{\emmg}{\emmee},
\label{eq:br}
\end{equation}
where $\mathcal{B}_{X}$ is the branching fraction of $\PGh\to X$; \Nmmee is the total number of signal events observed; \Nmmg is the number of reference channel events observed; and \emmee and \emmg are the experimental efficiencies of the $\PGmp\PGmm\Pep\Pem$ and $\PGmp\PGmm\PGg$ decay channels, respectively. 
The experimental efficiency for fully reconstructing an \PGh meson accounts for detector acceptance, trigger, and reconstruction requirements, and is estimated in Monte Carlo~(MC) simulations.
The numbers of signal and reference channel events are extracted from fits to the four-lepton invariant mass spectrum in each channel.
The spectrum for each decay mode, as well as the corresponding efficiencies, are derived from slightly different event selections that target each mode separately.
The efficiency ratio is corrected by scale factors derived from comparisons of data and MC simulation and applied where differences exist between the signal and reference channels.

The CMS apparatus~\cite{CMS:2008xjf,CMS:2023gfb} is a multipurpose, nearly hermetic detector, designed to trigger on~\cite{CMS:2020cmk,CMS:2016ngn,CMS:2024aqx} and identify electrons, muons, photons, and (charged and neutral) hadrons~\cite{CMS:2020uim,CMS:2018rym,CMS:2014pgm}. 
Its central feature is a superconducting solenoid of 6\unit{m} internal diameter, providing a magnetic field of 3.8\unit{T}. 
Within the solenoid volume are a silicon pixel and strip tracker, a lead tungstate crystal electromagnetic calorimeter (ECAL), and a brass and scintillator hadron calorimeter, each composed of a barrel and two endcap sections. 
Forward calorimeters extend the pseudorapidity coverage provided by the barrel and endcap detectors. 
Muons are reconstructed using gas-ionization detectors interleaved with the layers of the steel flux-return yoke outside the solenoid. 
More detailed descriptions of the CMS detector, together with a definition of the coordinate system used and the relevant kinematic variables, can be found in Refs.~\cite{CMS:2008xjf,CMS:2023gfb}. 
Tracks reconstructed in the tracker are combined with information from the ECAL and muon detectors to identify electron and muon candidates, respectively~\cite{CMS:2020uim,CMS:2018rym}. 
The electron momentum is estimated by combining the energy measurement in the ECAL, including the energy from bremsstrahlung photons, with the momentum measurement in the tracker. 
The muon momentum is that of the track. 
Conversions of photons in the beam pipe or detector are reconstructed by requiring two opposite-sign electrons to be successfully fitted to a common vertex with a constraint that both electrons are collinear at the vertex~\cite{CMS:2015myp}.

The CMS experiment uses a two-level trigger system~\cite{CMS:2016ngn}, with the first level~(L1) implemented in custom hardware~\cite{CMS:2020cmk}, and the high-level trigger~(HLT), in software~\cite{CMS:2024aqx}.
Most triggers in the standard HLT menu target events at high momenta, and thus employ high thresholds for transverse momentum \pt with respect to the beam axis (for example, 10--20\GeV for muons) to limit the overall event rate to a few kHz.
This rate is constrained by both the available storage and prompt-reconstruction resources.
However, CMS also has a ``data-parking" trigger path, whereby the running of the full event reconstruction can be delayed until the requisite computing resources are available~\cite{CMSParking}.
Data parking allows CMS to save lower-momentum events (for example, events with 3--4\GeV muons) without the risk of overloading the available offline computing resources.
For this analysis, a dimuon parking trigger with a peak rate of 1.6\unit{kHz} is used.
It is triggered by any two oppositely charged muons with \pt values measured at the HLT to be at least 4 and 3\GeV, with a common vertex of \chisq fit probability greater than 0.5\%. 
These muons must also have been reconstructed at the L1, which uses a simplified reconstruction algorithm to enable running on hardware.

Events accepted by the trigger are also required to satisfy several other criteria.
In both the signal channel and reference channel, events are required to have an opposite-sign muon pair and an opposite-sign electron pair such that a fit to a common vertex of the four leptons returns a common vertex of \chisq probability greater than 10\%. 
The vertexing is done using a Kalman filtering method~\cite{kalman}.
No missing electron track hits are allowed in the pixel detector, which comprises the innermost section of the tracker. 
This is to ensure that all photon conversions that are reconstructed as electron pairs occur either in the beam pipe or in the first layer of the pixel detector, limiting the systematic uncertainties related to the photon conversion probability. 
The HLT \pt requirements are reimposed offline, and electrons are required to have $\pt > 2\GeV$.
Muons should satisfy absolute pseudorapidity $|\eta| < 2.4$, while electrons must have $\abs{\eta}<1.44$ or $1.56<\abs{\eta}<2.5$, avoiding the transition region between the barrel and endcap ECAL detectors.

\begin{figure}[ht]
\centering
\includegraphics[width=0.5\textwidth]{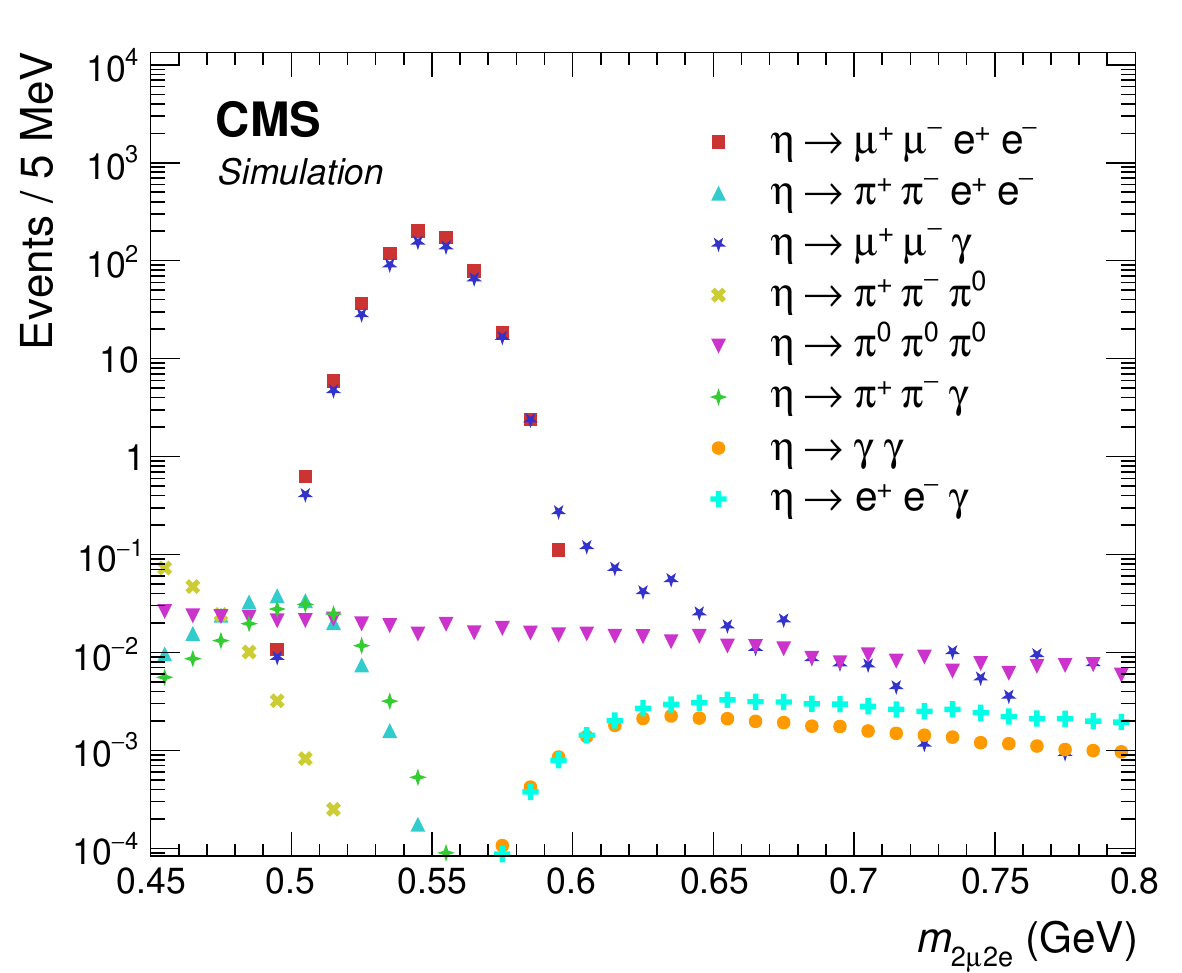}
\caption{Summary of expected background contributions to the yield measured in the signal channel based on a simplified MC simulation. The signal channel is shown in red squares, and is closely mimicked by the \twomuG channel, shown in blue stars. Uncertainties are not shown. The equivalent luminosity for the simulation is 38.0\fbinv.}
\label{fig:bkgs}   
\end{figure}

To determine if the data in the vicinity of the \PGh meson invariant mass peak represent solely \twomutwoe events or if it is contaminated with background, we use a simplified MC simulation to model the invariant mass spectra for various \PGh meson decays, including the signal process.
This study employs a simplified model for the physics of each process studied and does not include detector efficiency effects, following closely the strategy used in Ref.~\cite{cms4mu}.
Figure~\ref{fig:bkgs} shows the resulting invariant mass distributions for signal and the various potential backgrounds, which can resemble signal events because of misidentification of charged pions as muons, photon conversion to electron or muon pairs, or both.
All modes are weighted by the \PGh meson \pt-differential production cross section measured from \etatomumu decays, and normalized according to their world-average BFs~\cite{pdg} (except for the signal mode, which uses the BF predicted by Ref.~\cite{Escribano}), with corrections for misreconstruction effects, without further event selections. 

With one notable exception, potential backgrounds do not produce a significant number of events in the \PGh meson mass peak.
This is because assigning the muon mass to a pion track significantly shifts the reconstructed mass peak.
Moreover, this happens quite rarely in CMS, with roughly one in a thousand pions being misidentified as a muon.
Photon conversion to muons is also very rare, happening in the innermost tracker layer only once for every $10^{8}$ photons. 
The muons in such conversions receive a significant momentum kick from the recoiling nucleus, which shifts and flattens the apparent mass peak, as seen clearly in the $\twomuG$ distribution of Fig.~\ref{fig:bkgs}.
As a result of these effects, the only background with a significant probability of mimicking the \twomutwoe signal is the process \twomuG, with the photon converting to an electron pair in the innermost layer of the tracker.
The momentum kick received by the electrons resulting from a photon conversion to electrons is much smaller than the one to muons, and conversion to electrons is much more common, with around one in $10^{3}$ photons converting to an $\Pep\Pem$ pair in the innermost tracking layer. 
This decay is expected to contribute about half of the events in the \PGh meson signal mass peak (assuming the signal branching fraction prediction of Ref.~\cite{Escribano}), in the absence of cuts to mitigate it.
We will refer to this decay process, which is also the one used to provide the reference for the final BF measurement, as the resonant background.

Both electrons in the signal channel are required to pass an additional electron identification criterion.
The electron identification algorithm used is the boosted decision tree from Ref.~\cite{CMS:2020uim} that does not include isolation variables, since many \PGh mesons originate in jets.
The variables used for training this discriminator include reconstructed shower shape variables, energies, and track hit patterns.
While this requirement is already very effective at removing electrons originating from a photon conversion, we also require both electrons to not be part of any reconstructed conversion candidate.

The \twomuG process's ability to mimic the signal is also what makes it an ideal reference channel process. 
For the reference channel selection, we seek to minimize the efficiency of \twomutwoe events instead.
This is achieved by requiring both electrons to be part of a reconstructed conversion~\cite{CMS:2015myp}. 
The electron identification criterion used for the signal events is not applied, as this criterion would eliminate almost all reference channel events were it applied in this channel.

A full MC simulation is needed for the signal and resonant background processes to more accurately model the expected \PGh meson invariant mass shapes of both decay modes.
The \twomuG process is simulated with a \PYTHIA~8.3~\cite{pythia} ``particle gun" with an \PGh meson \pt and pseudorapidity distribution that covers the acceptance of the CMS detector, with the CP5~tune for the underlying event~\cite{cp5}.
The \PYTHIA particle gun uses a simple phase space model for the kinematic distributions of the decay products, rather than the more accurate VMD (as was used for the signal channel).
This phase space model does not accurately recreate the shape of the experimental dimuon invariant mass distribution found in Ref.~\cite{mmgObs}.
Therefore, a reweighting is applied to \twomuG simulated events to make the dimuon invariant mass squared distribution agree with the experimentally observed distribution of Ref.~\cite{mmgObs}, which was fitted to the theoretical model of Refs.~\cite{dalitz,krollWada}.
This is important, as the use of the reweighted distribution results in a 54\% larger \emmg (and hence, a 54\% larger measured $\mathcal{B}_{2\PGm 2\Pe}$) relative to the phase space distribution.
The products of additional proton-proton collisions occurring in the same or nearby bunch crossings are also included and the interaction of all generated particles with matter in the detector is simulated with \GEANTfour~\cite{geant}.

For the \twomutwoe signal process, a slightly modified version of the \textsc{Pluto} framework~\cite{pluto} is used to simulate the \PGh meson decay to a muon pair and an electron pair in its rest frame, using the VMD model. 
\textsc{Pluto} is a Monte Carlo event generator that specializes in low-energy processes like meson decays.
The only modification in this software was to adapt the \foure decay to \twomutwoe by replacing two of the electron masses with muon masses.
The resulting decay products are then interfaced with \PYTHIA to boost them into the laboratory frame.
The rest of the simulation is identical to that of the \twomuG process.
For both samples, the simulated events are weighted according to the \pt- and pseudorapidity-dependent \PGh meson production cross section. 
This cross section was determined using a sample of \etatomumu decays collected with the same trigger as for the signal and resonant background events and corrected for the effects of acceptance and efficiency using MC simulated events generated in a similar fashion as the other MC samples.
While the uncertainties in this cross section measurement are significant, they almost entirely cancel out in the final BF result because the ratio of the signal and reference channels is used, with the only remnant effect being the normalization of the small resonant background peak in the signal fit, which is accounted for using appropriate uncertainties.

Several corrections are made to the simulation to account for differences with data. 
The multiplicity of proton-proton interactions per bunch crossing in simulation is reweighted to match data.
Trigger efficiency corrections are derived by comparing \pt-dependent trigger efficiency distributions in simulation and data from monitoring triggers.
The reconstructed \pt of all simulated muons is multiplied by 0.9986, which is sufficient to align the mean reconstructed \etatomumu invariant mass to that in data to within 0.1\%. 
Following this correction, a similar correction of 1.001 is applied to the reconstructed \pt of all simulated electrons, which results in matching the mass of the signal channel to that in data.
A scale factor (SF) associated with the efficiency to identify an $\Pep\Pem$ pair is computed by comparing yields in data and simulation before and after applying the additional electron identification requirements in the signal channel, with a further requirement on the dielectron mass to eliminate the resonant background events.
The combined dielectron SF is found to be $0.60 \pm 0.08$, where the uncertainty includes the statistical uncertainty from the data and simulation as well as systematic uncertainties evaluated from varying the fitting functions used to describe the signal and background.

Including all corrections, the ratio of signal to reference channel efficiencies is found to be $\emmee/\emmg = 51.0 \pm 11.8$, where the uncertainty includes statistical and systematic components as detailed below.
This ratio is much lower than 1000 (the inverse of the fraction of photons that convert to electrons in the inner parts of the detector) because of the electron selection requirements, which are very different in the signal and reference channels. 

\begin{figure}[ht]
\centering
\includegraphics[width=0.5\textwidth]{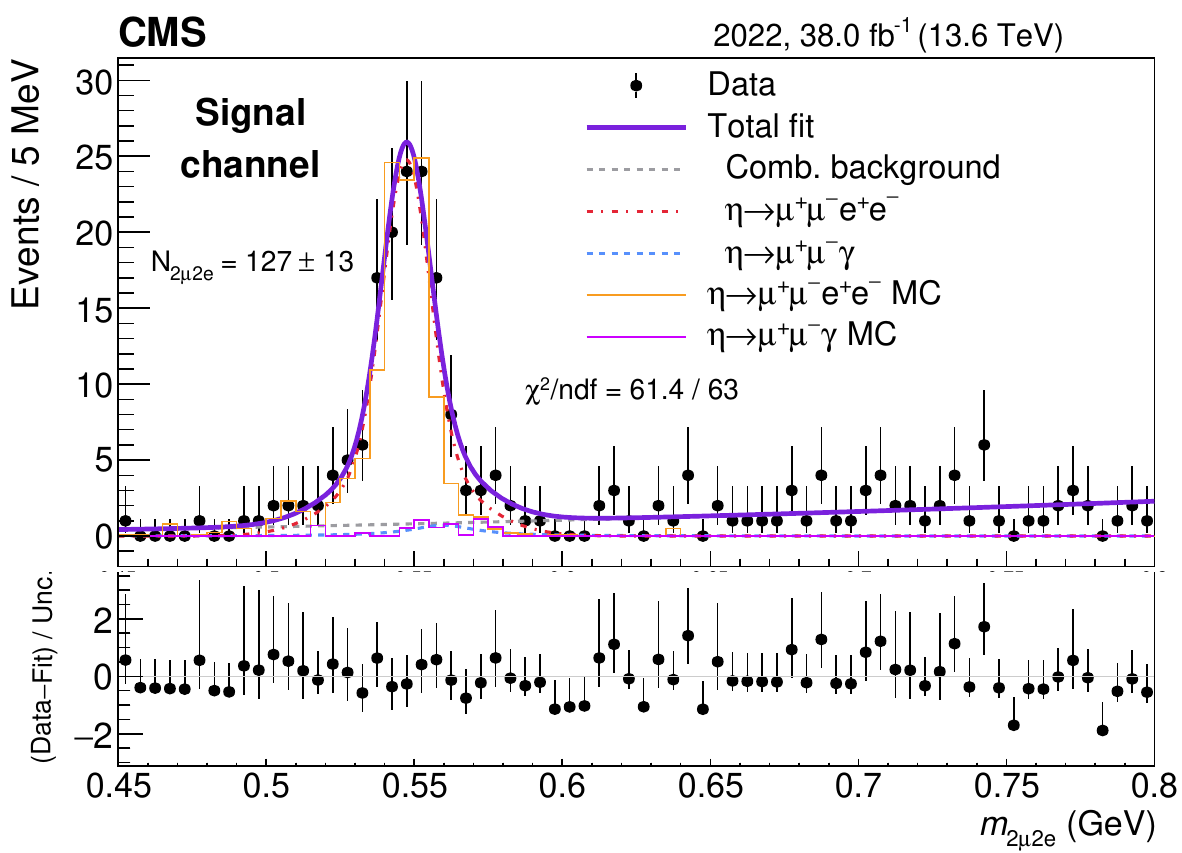}
\includegraphics[width=0.5\textwidth]{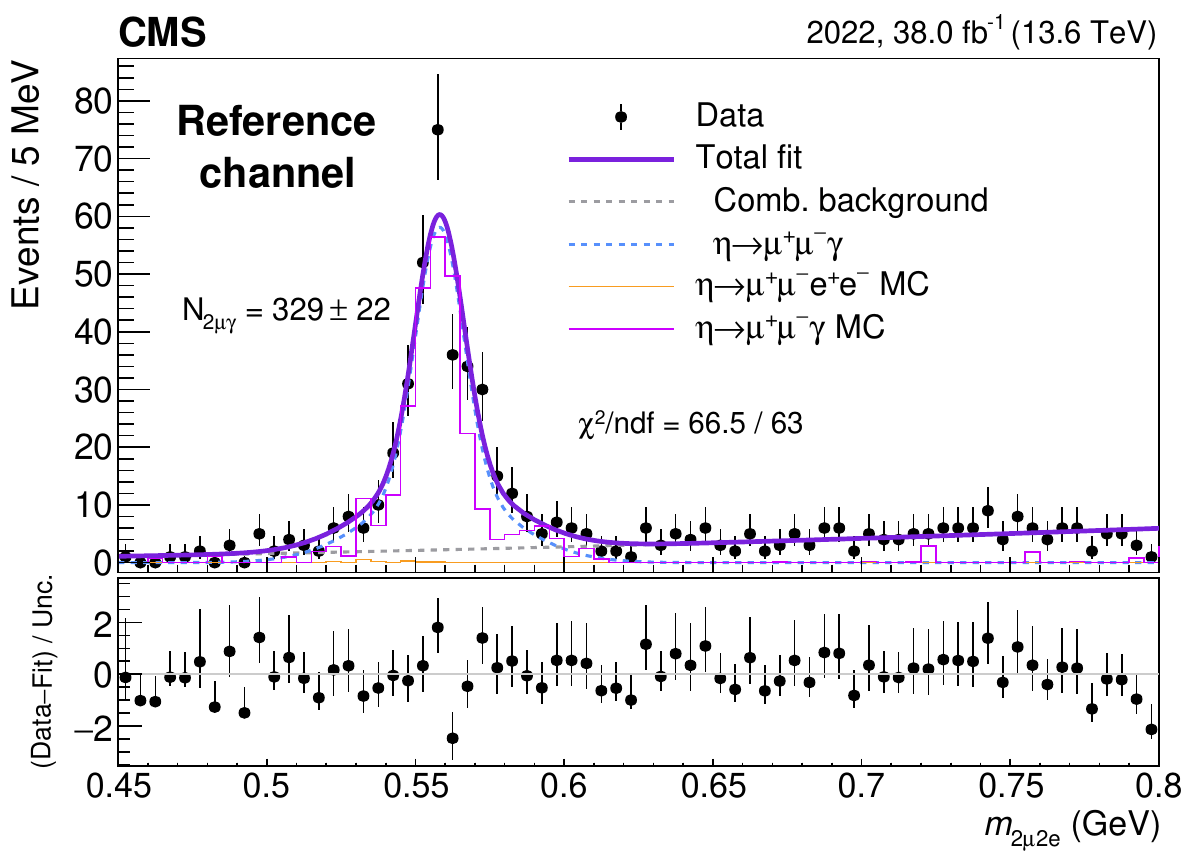}
\caption{Fit of the four-lepton invariant mass spectra in data for the signal (upper) and reference (lower) channels. Black points denote the data, the dashed grey curve is the combinatorial background fit to a threshold function, the cyan dashed curve is the \twomuG component of the fit, the red dot-dashed curve (signal channel only) is the \twomutwoe component of the fit, and the solid blue-violet curve is the overall fit including signal and background contributions. The orange distribution is the signal MC simulation (normalized to the BF prediction of $2.4\times10^{-6}$ by Ref.~\cite{Escribano}), and the magenta distribution is the \twomutwoe resonant background MC simulation (normalized to the world average BF measurement of $3.1 \times 10^{-4}$~\cite{pdg}), both also normalized using the approximate cross section derived from the \etatomumu channel. Points in the lower panel are pulls relative to the full fit model, (data $-$ fit)/$\sqrt{\smash[b]{\sigma_{\textrm{data}}^2 - \sigma_{\textrm{fit}}^2}}$.}
\label{fig:datafit}
\end{figure}

To estimate \Nmmee, the invariant mass distribution of the signal channel is fit with a binned extended maximum likelihood fit.
For signal, a double-Gaussian model is used, while for resonant background, a Breit--Wigner model is used. 
These empirically motivated models are chosen because they provide fits with the best \chisq values, and the fitted yields are found roughly in the middle of the range obtained with other models (such as combinations of Gaussians and Crystal-Ball functions~\cite{CB}) exhibiting similar \chisq values.
A double-Gaussian signal MC fit is used as a starting point for the fit to data, with the normalization floating freely and other fit parameters constrained by the uncertainties from the MC fit, but with relaxed bounds if the best fit value is found to be near a boundary.
All parameters of the resonant background Breit--Wigner function found from a fit to the MC spectrum are fixed in the final fit to data, with the normalization derived from the measured cross section, the world-average value of $\BtwomuG / \mathcal{B}(\PGh \to \PGmp \PGmm)$, and the efficiency from simulation for \twomuG events to pass the signal channel criteria.
In addition to the double-Gaussian signal and Breit--Wigner resonant background, a small continuous background component is added, and is modeled with a threshold function, proportional to $[m_{2\PGm 2\Pe} - (2m_{\Pe} + 2m_{\PGm})]^\beta$, where $m_{\Pe}$ is the electron mass, $m_{\PGm}$ is the muon mass, and $\beta$ is the fit parameter.
The fit to data, along with the signal and resonant background MC distributions, is shown in Fig.~\ref{fig:datafit} (upper).

The central value of the resonant background invariant mass distribution, at 555\MeV, is significantly higher than the world average value of the \PGh meson mass, 548\MeV~\cite{pdg}.
This is a consequence of these electrons being produced in the beam pipe or detector rather than from the proton-proton collision vertex.
This effect is also seen in the reference channel.
The fitted signal yield is $\Nmmee = 127 \pm 13$, on top of about 4 resonant background events and about 10 combinatorial background events in the signal window (0.52--0.58\GeV for the signal channel).
Comparing the \chisq values for the signal+background and background-only models via the likelihood-ratio test, the resulting test statistic corresponds to a significance far exceeding 5 standard deviations against the \mbox{background-only} hypothesis.

The fit to the data observed in the reference channel is shown in Fig.~\ref{fig:datafit} (lower).
As in the signal channel, the model for the peak is a double-Gaussian function, and the combinatorial background is modeled by the threshold function.
The starting parameters for the fit are taken from fits to the MC invariant mass distribution and are allowed to float in the fit to data, just as in the signal channel fit.
In this channel, the \twomutwoe signal mode is strongly suppressed by the inverted conversion veto, as seen from the orange MC distribution.
This fit result is $\Nmmg = 329 \pm 22$.

Several sources of systematic uncertainty affect the result, while many others cancel when taking the ratio to the reference channel.
The fit model uncertainties are estimated by redoing the fits to each channel with two alternative models that give similar goodness-of-fit metrics. 
The alternative signal models used are the Crystal-Ball plus Gaussian model and the Voigtian model, and the alternative combinatorial background models used are first- and second-order polynomials. 
For each of the four sources (signal and background function for signal and reference channels), the maximum variation is taken as the systematic uncertainty. 
An additional 1\% uncertainty arises from a 27\% uncertainty in the normalization of the resonant background in the signal channel fit, which comes from uncertainties in our measured \etatomumu cross section and the world-average $\BtwomuG / \mathcal{B}(\PGh \to \PGmp \PGmm)$ branching ratio. 
Summing the five uncertainties in quadrature results in a 12\% signal yield systematic uncertainty.
The uncertainty related to the conversion veto requirement efficiency for the reference channel is important to evaluate as it is a significant difference between the two channels.
This uncertainty is estimated by combining data-to-simulation checks from three different sources. 
The first contribution is obtained from data-simulation comparisons in the \twomuG channel invariant mass spectrum, without the additional electron identification required, with two different selection criteria: with, or without, a photon conversion veto required. 
The second contribution is from a study of the tracker material that found an agreement better than 4\% between data and simulation for the amount of material in the beam pipe and first pixel layer.
Lastly, the uncertainty in the track reconstruction efficiency for decay vertices occurring in the first pixel layer is evaluated using $\PKzS\to\PGpp\PGpm$ decays, as described in Ref.~\cite{displaced}.
Altogether, the uncertainty in the conversion requirement efficiency amounts to 13\% (including 5\% from the data-simulation comparison, 4\% from the material budget uncertainty, and 11\% from the track reconstruction efficiency study).
Another requirement that can affect the signal and reference channels differently is the vertex probability requirement. 
Removing this cut from both channels for the data and from the corresponding simulated events used to calculate the efficiency results in an 8\% change in the branching ratio, which is included as a systematic uncertainty.
The uncertainty in the kinematic distribution in the reference channel is found by reweighting the simulated events such that the form factor in the theoretical model is 1 standard deviation above or below the central value reported in Ref.~\cite{mmgObs}.
The maximum deviation in the calculated efficiency is taken, which amounts to 3\%.
In the signal channel, event reweighting is done to make both the dimuon and dielectron invariant mass distributions approximately agree with the theoretical predictions of Ref.~\cite{Escribano}, and the change in the efficiency is taken as the uncertainty (11\%).
For the signal channel, unlike the separate reweighting performed in the reference channel, this reweighting is done only to estimate the uncertainty and not as a correction on the result (which would have resulted in the measured BF being 11\% lower). 
Combined with the 14\% uncertainty in the dielectron identification efficiency SF and 8\% from the limited number of simulated events, these uncertainties add up to a total systematic uncertainty of 27\%. 

Using the yields and efficiencies defined in Eq.~(\ref{eq:br}), the measured BF ratio is
\ifthenelse{\boolean{cms@external}}
{ 
\begin{align*}
    \frac{\mathcal{B}_{2\PGm 2\Pe}}{\mathcal{B}_{2\PGm\PGg}} &= [7.3 \pm 0.9 \stat \pm 2.0 \syst]\times 10^{-3} \\
                                                             &= [7.3 \pm 2.2]\times 10^{-3}.
\end{align*}
} 
{ 
\begin{equation*}
    \frac{\mathcal{B}_{2\PGm 2\Pe}}{\mathcal{B}_{2\PGm\PGg}} = [7.3 \pm 0.9 \stat \pm 2.0 \syst]\times 10^{-3} = [7.3 \pm 2.2]\times 10^{-3}.
\end{equation*}
} 
Using the world average value of $\mathcal{B}_{2\PGm\PGg} = (3.1 \pm 0.4) \times 10^{-4}$~\cite{pdg},
\ifthenelse{\boolean{cms@external}}
{ 
\begin{align*}
    \mathcal{B}_{2\PGm 2\Pe} &= [2.4 \pm 0.3 \stat \pm 0.6 \syst \pm 0.3 ( \mathcal{B}_{2\PGm\PGg}) ] \times 10^{-6} \\
                             &= [2.4 \pm 0.8] \times 10^{-6}.
\end{align*}
} 
{ 
\begin{equation*}
    \mathcal{B}_{2\PGm 2\Pe} = [2.4 \pm 0.3 \stat \pm 0.6 \syst \pm 0.3 ( \mathcal{B}_{2\PGm\PGg}) ] \times 10^{-6} = [2.4 \pm 0.8] \times 10^{-6}.
\end{equation*}
} 
The previous upper limit for this BF was $6.88 \times 10^{-6}$ at 90\% confidence level, set by the BESIII experiment~\cite{besIII}.
Our measurement is consistent with the latest standard model predictions of $2.3\text{--}2.4 \times 10^{-6}$, based on data-driven approaches~\cite{Kampf,Escribano}, while other models yield predictions as low as $0.6 \times 10^{-6}$~\cite{Lih} and as high as $5 \times 10^{-6}$~\cite{Bijnens}.

In summary, the decay \twomutwoe has been observed for the first time, with a significance much larger than 5 standard deviations, with data collected by the CMS experiment at the LHC in 2022. 
The branching fraction of this decay is measured with respect to the \twomuG decay mode: $\Btwomutwoe / \BtwomuG = (7.3 \pm 2.2)\times 10^{-3}$.
Incorporating the world-average value of \BtwomuG results in an \twomutwoe branching fraction of $(2.4 \pm 0.8) \times 10^{-6}$, which is about a factor of three below the previous upper limit and is consistent with the latest standard model predictions.
This measurement provides useful information for the determination of transition form factors that are used for theoretical calculations, such as the muon anomalous magnetic moment.

\begin{acknowledgments}
We congratulate our colleagues in the CERN accelerator departments for the excellent performance of the LHC and thank the technical and administrative staffs at CERN and at other CMS institutes for their contributions to the success of the CMS effort. In addition, we gratefully acknowledge the computing centers and personnel of the Worldwide LHC Computing Grid and other centers for delivering so effectively the computing infrastructure essential to our analyses. Finally, we acknowledge the enduring support for the construction and operation of the LHC, the CMS detector, and the supporting computing infrastructure provided by the following funding agencies: SC (Armenia), BMBWF and FWF (Austria); FNRS and FWO (Belgium); CNPq, CAPES, FAPERJ, FAPERGS, and FAPESP (Brazil); MES and BNSF (Bulgaria); CERN; CAS, MoST, and NSFC (China); MINCIENCIAS (Colombia); MSES and CSF (Croatia); RIF (Cyprus); SENESCYT (Ecuador); ERC PRG and PSG, TARISTU24-TK10 and MoER TK202 (Estonia); Academy of Finland, MEC, and HIP (Finland); CEA and CNRS/IN2P3 (France); SRNSF (Georgia); BMFTR, DFG, and HGF (Germany); GSRI (Greece); MATE and NKFIH (Hungary); DAE and DST (India); IPM (Iran); SFI (Ireland); INFN (Italy); MSIT and NRF (Republic of Korea); MES (Latvia); LMTLT (Lithuania); MOE and UM (Malaysia); BUAP, CINVESTAV, CONACYT, LNS, SEP, and UASLP-FAI (Mexico); MOS (Montenegro); MBIE (New Zealand); PAEC (Pakistan); MSHE, NSC, and NAWA (Poland); FCT (Portugal); MESTD (Serbia); MICIU/AEI and PCTI (Spain); MOSTR (Sri Lanka); Swiss Funding Agencies (Switzerland); MST (Taipei); MHESI (Thailand); TUBITAK and TENMAK (T\"{u}rkiye); NASU (Ukraine); STFC (United Kingdom); DOE and NSF (USA).
\end{acknowledgments}

\bibliography{auto_generated}
\cleardoublepage \appendix\section{The CMS Collaboration \label{app:collab}}\begin{sloppypar}\hyphenpenalty=5000\widowpenalty=500\clubpenalty=5000\cmsinstitute{Yerevan Physics Institute, Yerevan, Armenia}
{\tolerance=6000
A.~Hayrapetyan, V.~Makarenko\cmsorcid{0000-0002-8406-8605}, A.~Tumasyan\cmsAuthorMark{1}\cmsorcid{0009-0000-0684-6742}
\par}
\cmsinstitute{Institut f\"{u}r Hochenergiephysik, Vienna, Austria}
{\tolerance=6000
W.~Adam\cmsorcid{0000-0001-9099-4341}, L.~Benato\cmsorcid{0000-0001-5135-7489}, T.~Bergauer\cmsorcid{0000-0002-5786-0293}, M.~Dragicevic\cmsorcid{0000-0003-1967-6783}, P.S.~Hussain\cmsorcid{0000-0002-4825-5278}, M.~Jeitler\cmsAuthorMark{2}\cmsorcid{0000-0002-5141-9560}, N.~Krammer\cmsorcid{0000-0002-0548-0985}, A.~Li\cmsorcid{0000-0002-4547-116X}, D.~Liko\cmsorcid{0000-0002-3380-473X}, M.~Matthewman, J.~Schieck\cmsAuthorMark{2}\cmsorcid{0000-0002-1058-8093}, R.~Sch\"{o}fbeck\cmsAuthorMark{2}\cmsorcid{0000-0002-2332-8784}, M.~Shooshtari\cmsorcid{0009-0004-8882-4887}, M.~Sonawane\cmsorcid{0000-0003-0510-7010}, W.~Waltenberger\cmsorcid{0000-0002-6215-7228}, C.E.~Wulz\cmsAuthorMark{2}\cmsorcid{0000-0001-9226-5812}
% ...
\par}
$^{1}$Also at Yerevan State University, Yerevan, Armenia\\
$^{2}$Also at Technische Universit\"{a}t Wien, Vienna, Austria\\
% ...
\end{sloppypar}
\end{document}